\documentclass[conference, 10pt]{IEEEtran} 

\usepackage[utf8]{inputenc} % allow utf-8 input
\usepackage[T1]{fontenc}    % use 8-bit T1 fonts
\usepackage{booktabs}       % professional-qualitytables
\usepackage{graphicx}
\usepackage{nicefrac}       % compact symbols for 1/2, etc.
\usepackage{microtype}      % microtypography
\usepackage[cmex10]{amsmath}
\usepackage{amssymb}
\usepackage{amsfonts}
\usepackage{amsthm}
\usepackage{thmtools}
\usepackage{url}
\usepackage{cite}
\usepackage[hang]{subfigure}
\usepackage{tikz}
\usepackage[linesnumbered,ruled]{algorithm2e}

\makeatletter
\newcommand{\removelatexerror}{\let\@latex@error\@gobble}
\makeatother
\usepackage{color}

\definecolor{Rafael}{RGB}{139,0,0} % You can pick a different color if you like ;D

\definecolor{rick}{RGB}{0,139,0}

\definecolor{old}{RGB}{0,0,139}

\title{Reinforce Security: A Model-Free Approach Towards Secure Wiretap Coding}
\author{
	  \IEEEauthorblockN{Rick Fritschek$^{\ast}$, Rafael F. Schaefer$^{\dagger}$, and Gerhard Wunder$^{\ast}$\\[2mm]}
	  \IEEEauthorblockA{\small
	    \begin{tabular}{cc}
	       \begin{tabular}{c}
	           $^{\ast}$ Heisenberg Communications and Information Theory Group\\
                        Freie Universit\"at Berlin \\
                        \texttt{\{rick.fritschek, g.wunder\}@fu-berlin.de}
	       \end{tabular}
	       \begin{tabular}{c}
	           $^{\dagger}$ Lehrstuhl f\"ur Nachrichtentechnik/Kryptographie und Sicherheit\\
	                        Universit{\"a}t Siegen \\
                            \texttt{rafael.schaefer@uni-siegen.de}
	       \end{tabular}
	  \end{tabular}
}
\thanks{This work was supported by the German Research Foundation (DFG)
under Grants FR 4209/1-1 and SCHA 1944/7-1.}}
\IEEEoverridecommandlockouts
\begin{document}

\maketitle

\begin{abstract}
The use of deep learning-based techniques for approximating secure encoding functions has attracted considerable interest in wireless communications due to impressive results obtained for general coding and decoding tasks for wireless communication systems. Of particular importance is the development of model-free techniques that work without knowledge about the underlying channel. Such techniques utilize for example generative adversarial networks to estimate and model the conditional channel distribution, mutual information estimation as a reward function, or reinforcement learning. In this paper, the approach of reinforcement learning is studied and, in particular, the policy gradient method for a model-free approach of neural network-based secure encoding is investigated.  Previously developed techniques for enforcing a certain co-set structure on the encoding process can be combined with recent reinforcement learning approaches. This new approach is evaluated by extensive simulations, and it is demonstrated that the resulting decoding performance of an eavesdropper is capped at a certain error level.
\end{abstract}
\section{Introduction}
\label{sec:intro}
A recent breakthrough in wireless communication is the deep learning-based approximation of encoding and decoding functions. These deep learning approaches are based on neural network (NN) representations of these functions, where the weights are optimized to yield encoder-decoder pairs for reliable communication over noisy channels. In particular, one of the first approaches looked at end-to-end learning of these communication systems, by utilizing the so-called autoencoder approach \cite{OShea2017}. This approach demonstrated that the resulting NN-based encoder and decoder can perform close to classical baseline techniques \cite{SBrink2018}. These approaches usually utilize an optimization over a minimum squared error term or a cross-entropy loss term via variants of gradient descent. There, the loss function is linked to the decoder, and it is not possible to train the encoder without it. Another line of work optimized the encoder without a corresponding decoder by optimizing a mutual information approximation over samples of the channel input and output \cite{Fritschek_SPAWC19}, \cite{Fritschek_SPAWC20}. Furthermore, a series of recent works has adapted these NN-based encoder-decoder pairs for reliable and secure communication, i.e., to learn secure encoding functions by introducing a secrecy constraint into the optimization. In wireless communication, and in particular information theory, secrecy means to bound the information leakage, i.e., information about a confidential message that is leaked to unintended receivers (eavesdroppers). In general, it is hard to compute or even approximate the leakage if one has only access to samples, as the the leakage is defined by mutual information. This makes it hard to straightforwardly optimize an NN encoder-decoder pair with a secrecy constraint. Recent examples that try to tackle this problem are: In \cite{Karl_ICASSP} the leakage is approximated by tracking the NN, which has the drawback that it cannot use stochastic gradient descent. Another example is \cite{zhang2019deep}, where NNs were utilized to learn appropriate precoding for a MIMO Gaussian wiretap scenario. In \cite{ICC_DL_Sec}, a secrecy constraint was introduced by altering the one-hot representation of the input of a structure enforcing decoder, where the resulting secrecy enabling loss function is based on the standard cross-entropy loss. This cross-entropy loss will impose a clustering in the transmit constellations and, accordingly, will imitate the classical co-set coding approach for security. 

A third branch within this recent deep learning-assisted wireless communication field is to make these methods channel model independent. Some of the recent works include the use of 
\emph{i) Generative adversarial networks (GANs):} GANs were introduced in \cite{goodfellow2014generative} and are composed of a generative NN and a discriminative NN. The generative NN gets a noise input and has the goal to generate a certain wanted distribution. The discriminator, on the other hand, has as two inputs, samples from the generated distribution and samples from the real distribution, with the goal to distinguish between fake and real samples. Both NNs are now alternatingly optimized, and the resulting generative NN can be used as an approximation of the real distribution. In the context of wireless communication, one can approximate the channel distribution by samples and use the generator as a piece within the NN encoder-decoder chain \cite{Cond_GAN_AE,oShea2018GAN}.  This was recently utilized to enable a form of secure communication in \cite{Gunduz_GAN_sec}.
\emph{ii) Mutual information estimators:} A recent breakthrough has shown that mutual information can be approximated through sampling of the random variables with the help of NNs \cite{belghazi2018mine}. This was utilized in \cite{Fritschek_SPAWC19} to estimate the mutual information between channel input and output samples and use this as a metric to train the NN encoder to maximize mutual information. This approach has the advantage that it tackles the communication problem from the information-theoretic foundations. However, these mutual information approximations are lower bounds and cannot be used as approximations for the leakage as an upper bound would be needed. A possible workaround for certain channel is given by \cite{molavipour2020conditional} which shows how a conditional mutual information can be estimated. 
\emph{iii) Reinforcement learning (RL)}: The third line of work in this branch is to utilize policy gradient methods. Considering the messages and codewords of the communication system as states and actions, and integrating the channel and decoding function into the reward function evaluation, one can formulate a corresponding policy gradient problem where only the reward function evaluations are used, not its derivatives. The idea to utilize RL to learn optimized NN encoder-decoder pairs was introduced in \cite{Aoudia-RL}, and subsequently extended to noisy feedback links in \cite{goutay2018deep}. The disadvantage of using model-free reinforcement learning is that the approach is necessarily less effective than utilizing more structure, i.e, gradients of the channel function \cite{recht2019tour}. However, its generality and the ease of implementation makes it an option worth to be further explored.

Our contribution is now that we combine the previously mentioned RL-based learning approach that uses the policy gradient theorem with the secure encoding approach that uses the altered one-hot input distribution for model-aware training. The combination of them is particularly promising, because the one-hot security approach utilizes a cross-entropy based metric. This makes it possible to construct a novel per sample loss, which conserves the structure enforcing properties of the secure encoding approach. With this, we show how to build a secrecy enabling per sample loss for encoding and demonstrate that the resulting method can learn to cluster the codewords into co-sets and therefore enable secrecy with appropriate encoding without model knowledge.

{\bf Notation:} We stick to the convention of upper case random variables $X$ and lower case realizations $x$, i.e., $X\sim p(x)$, where $p(x)$ is the probability mass or density function of $X$. Moreover, $p(\mathbf{x})$ is the probability mass or density function of the random vector $\mathbf{X}$. We also use $|\mathcal{X}|$ to denote the cardinality of a set $\mathcal{X}$. The expectation is denoted by $\mathbb{E}[\cdot]$.

\section{Wiretap Channel}
\label{InfoThModel}

\begin{figure}
    \centering
    \includegraphics[scale=0.73]{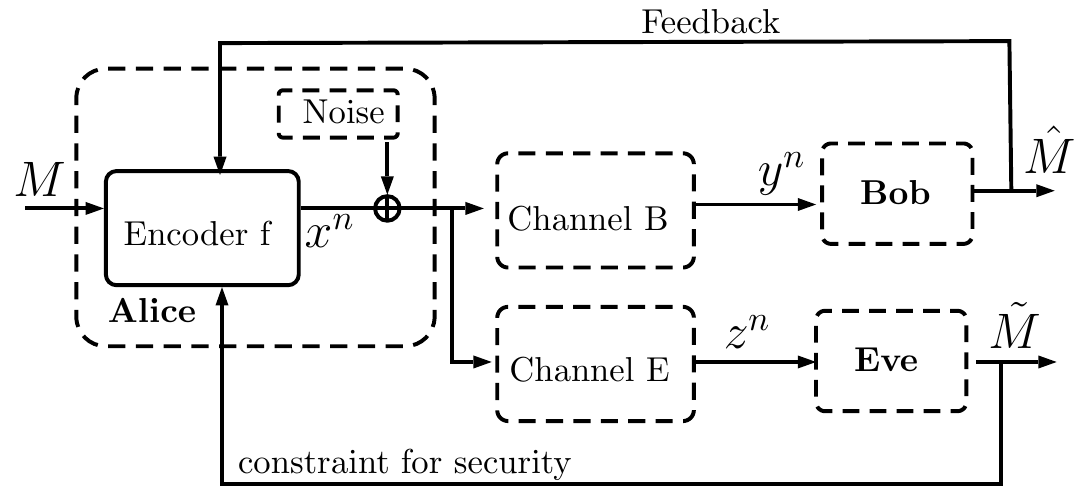}
    \caption{The wiretap channel. The encoder $f$ of Alice is trained to enable secure communication to Bob. Alice's signal will be perturbed, to enable exploration for the policy gradient method. Moreover, a security constraint is included by invoking an exemplary eavesdropper Eve. After that the channel to Bob estimates a per sample loss, which is designed to allow secure encoding and will be fed back to the encoder. The encoder can now use the policy gradient theorem to train on the security enabling per sample loss, without channel knowledge.}
    \label{fig:my_label}
\end{figure}

%\begin{figure}
%\centering
%\includegraphics[scale=0.8]{G-WTC-D2.pdf}
%\caption{Degraded Gaussian wiretap channel. The confidential communication is %between Alice and Bob, while Eve tries to eavesdrop upon it.}
%\label{Wiretap-degraded-classical}
%\end{figure}
In this paper, we consider the communication scenario with a transmitter Alice, a legitimate receiver Bob, and an eavesdropper Eve as shown in Fig.~\ref{fig:my_label}. Alice wants to transmit a confidential message $m\!\in\! \mathcal{M}\!=\!\{1,2, \ldots, 2^{nR}\}$ with rate $R$ by using an encoding function $f$ that encodes the message $m$ into a codeword $\mathbf{x}(m)\in\mathbb{C}^n$ and transmit it over the noisy channel to Bob who needs to decode the message. At the same time, Eve needs to be kept ignorant about the message. This model is called the \emph{wiretap channel} and provides the basic scenario to investigate at what rate messages can be reliably sent to a legitimate receiver (Bob) while providing secrecy against a wiretapper (Eve).

To this end, for every message $m\in\mathcal{M}$, we assume an average power constraint $\tfrac{1}{n}\sum_{i=1}^n|x_i(m)|^2\leq P$ on the corresponding codewords $\mathbf{x}(m)$. The channel from Alice to Bob is given by the transition probability density $p_{\mathbf{Y}|\mathbf{X}}(\mathbf{y}|\mathbf{x})$ for input and output sequences $\mathbf{x}$ and $\mathbf{y}$. If the channel is further memoryless, one has 
$
   p_{\mathbf{Y}|\mathbf{X}}(\mathbf{y}|\mathbf{x})=\prod_{i=1}^{n} p(y_i|x_i)$, i.e., the output at time instant $i$ depends only on the corresponding input at time instant $i$ and is independent of the previous inputs. The channel from Alice to Eve can be defined accordingly.
The receiver Bob uses a decoder $g$ to estimate a message  $g(\mathbf{y})=\hat{m}$ which should recover the original message. Moreover, the block error rate $P_e$ is defined as the average probability of error over all messages
\begin{equation}
    P_e= \frac{1}{|\mathcal{M}|}\sum_{m\in\mathcal{M}} \mbox{Pr}(\hat{M}\neq m | M = m).
\end{equation}

Without any secrecy constraint, the maximal transmission rate $R$ such that the error $P_e$ vanishes for sufficiently large $n$ is called the \emph{capacity} $C$ of the channel and is known to be
%The general communication problem is now to find the maximal communication rate $R$, such that the error $P_e$ can be made arbitrary small for a sufficiently large $n$. This optimal rate $R$ is called the capacity of the channel and is known to be 
$C=\max_{p(x)} I(X;Y)$ for discrete memoryless channels, cf. for example \cite{CoverThomas06ElementsInformationTheory}.

Since Eve is eavesdropping upon the legitimate communication, we impose a secrecy constraint to keep the transmitted message confidential. Usually, information theoretic principles are invoked and the information leakage to the eavesdropper is required to vanish. There are multiple definitions including weak\cite{Wyner75} or strong secrecy\cite{maurer2000strong}. For strong secrecy, this criterion is defined as
\begin{equation}
\lim_{n \rightarrow \infty} I(M;\mathbf{Z})=0
\label{screcy criterion}
\end{equation}
with $\mathbf{Z}=(Z_1,Z_2,...,Z_n)$ the channel output at Eve, cf. for example \cite{BlochBarros-2011-PhysicalLayerSecurity}.

Now, the \emph{secrecy capacity} characterizes the maximum transmission rate $R$ at which Bob can reliably decode the transmitted message while, simultaneously, Eve is kept in the dark, i.e., the secrecy criterion \eqref{screcy criterion} is satisfied. As we pointed out above, it is unfortunately, still a major challenge to optimize NNs according to such a constraint, as it is challenging to estimate an upper bound for the mutual information from samples. Instead, as in \cite{ICC_DL_Sec}, we opt for a secrecy criterion based on the cross-entropy metric. This goes well with the overall approach of using NNs to approximate optimal encoder-decoder pairs, as well as the reinforcement learning-based technique of policy gradient, because it can be implemented on a per sample basis.

\section{Reinforcement Learning for Wireless Communications}
\label{sec:RL-WC}

The goal of reinforcement learning is to optimize a reward $r(s_i,a_i)$ based on states $s_i$ of an environment and actions $a_i$ taken in this environment. The actions $a_i$ are done by a policy $\pi_{\theta}(a_i|s_i)$, parameterized by $\theta$, based on the state $s_i$. One can then write down the probability of the whole trajectory $\tau$ of actions and states as $p_{\theta}(s_1,a_1,\ldots)=p_{\theta}(\tau)=p(s_1)\prod_i \pi_{\theta}(a_i|s_i)p(s_{i+1}|s_i,a_i)$ and state the optimization problem as
\begin{equation*}
    \max_{\theta} J(\theta) = \max_{\theta}  \: \mathbb{E}_{\tau\sim p_{\theta}(\tau)} [r(\tau)].
\end{equation*}
Since we want to maximize $J(\theta)$, we can use gradient ascent. The gradient can be written as 
%\begin{IEEEeqnarray}{rCl}
%    \nabla_{\theta}J(\theta)&=&\nabla_{\theta} \int p_{\theta}(\tau) r(\tau) \text{d}\tau\IEEEnonumber\\
%   &=& \int r(\tau) \nabla_{\theta} p_{\theta}(\tau) \text{d}\tau\IEEEnonumber\\
%   &=& \int r(\tau) p_{\theta}(\tau)\nabla_{\theta} \log p_{\theta}(\tau) \text{d}\tau\\
%   &=& \mathbb{E}_{p_{\theta}(\tau)} \left[ r(\tau) \nabla_{\theta} \log p_{\theta}(\tau) \right]\IEEEnonumber.
%\end{IEEEeqnarray}
%Note, that a rigorous derivation involves unrolling of the trajectories, details can be seen in \cite{sutton2018reinforcement}. Here, we only use the identity $\nabla f(x)=f(x)\tfrac{\nabla f(x)}{f(x)}=f(x)\nabla \log f(x)$ in step ($5$), called the log-likelihood trick. Now, the expectation can be approximated by a sample average. Moreover, note that 
%\begin{equation}
%   \nabla_{\theta} \log p_{\theta}(\tau) = \sum_i \nabla_{\theta} \log \pi_{\theta}(a_i|s_i) 
%\end{equation} and we can therefore state the finite case as 
\begin{equation*}
    \nabla_{\theta} J(\theta) = \mathbb{E}_{p_{\theta}} \left[ \sum_i r(s_i,a_i) \nabla_{\theta} \log \pi_{\theta}(a_i|s_i) \right],
\end{equation*} where we see that we do not need the derivative of $p(s_{i+1}|s_i,a_i)$, but only the derivative of the policy $\pi_{\theta}(a_i|s_i)$ is needed. A complete derivation of this result can be found for example in \cite{sutton2018reinforcement}. The optimization step is therefore free of any model knowledge. This is a very convenient form because the gradient now involves only the reward function evaluations and not its derivatives. Note that this only works for stochastic policies, otherwise system model knowledge is required. 

For the particular case of wireless communications, one can identify messages as states, and the sent codewords as actions. In standard wireless encoding situations, one has deterministic encoders and therefore a deterministic policy. However, \cite{Aoudia-RL} showed that one can add noise on the codeword $\mathbf{x}$, arriving at a perturbed version $\mathbf{x}_p = \mathbf{x}+\mathbf{w}$, where $\mathbf{w}$ can be zero mean Gaussian noise. The policy therefore takes in a message $m$, and outputs a perturbed codeword $\mathbf{x}_{p}$. Denoting the decoder as $g$, which has a softmax output, the channel function as $h$, and assuming one-hot input, one can define a per sample reward as $r_i:=\log g(h(\mathbf{x}_{p,i}))$, which gives the cross-entropy over the sent messages with the estimated messages. One can therefore optimize the encoder, via $\theta$, without having access to the derivatives of the channel or the decoder. However, note that the decoder is an integral part of the per sample loss, which is why encoder and decoder need to be trained alternatingly.

\section{Encoding-Decoding Procedure and Implementation}

\label{sec:Enc-Dec-Impl}
The encoder of Alice is modelled as a NN with weights $\theta$, one fully-connected hidden layer with an elu activation function, and a linear output layer. The encoder gets one-hot encoded messages, i.e., binary vectors $\mathbf{m}_{\text{oh}}\in\mathbb{F}_2^ {|\mathcal{M}|}$ of the form $(0,...,0,1,0,...,0)$ which have a one at the $i$-th position, representing the $i$-th message of $\mathcal{M}=\{1,...,|\mathcal{M}|\}$. The output of the network is then normalized to have unit power and is shaped from $2n$ real values to $n$ complex values or codewords $\mathbf{x}_{\theta}(m)$, which are sent over the legitimate channel. 

The decoder $g$ of Bob, which receives the noisy channel output $\mathbf{y}$, is also modelled as a NN with weights $\psi$, with one fully-connected hidden layer with elu activation. Moreover, it has a softmax output layer, which gives an estimate of the probability distribution of the sent message. Let $\mathbf{\nu}\in \mathbb{R}^{|\mathcal{M}|}$ be the output of the last dense layer in the decoder network. The softmax function takes $\mathbf{\nu}$ and returns a vector of probabilities for the message set, i.e., $\mathbf{p}\in (0,1)^{|\mathcal{M}|}$, where the entries $p_m$ are calculated by
\begin{equation*}
    p_m=f(\mathbf{\nu})_m:=\frac{\exp(\nu_m)}{\sum_i \exp(\nu_i)}.%\quad \mbox{for } m \in \{1,\ldots, |\mathcal{M}|\}.
\end{equation*}
The decoder then declares the estimated message to be $\hat{m}=\arg\max_m p_m$. Furthermore, it outputs the estimated probabilities $p_m$ of the received message index and feeds it into a cross-entropy function together with the true index $m$
\begin{IEEEeqnarray*}{rCl}
H(M,\hat{M})&=&-\sum_{m\in \mathcal{M}}p(m)\log p_{\text{decoder}}(m)\IEEEnonumber\\
&=&-\mathbb{E}_{p(m)}[\log  p_{\text{decoder}}(m)],
\end{IEEEeqnarray*} which is estimated by averaging over the sample size $k$, which yields the cross-entropy cost function to optimize the decoder weights~$\psi$
\begin{IEEEeqnarray}{rCl}
    L(\psi)&=&-\frac{1}{k}\sum_{i=1}^k \log p_{m_i},
    \label{Loss_unsecure}
\end{IEEEeqnarray} where $m_i$ represents the index of the message of the $i$-th sample. The per sample loss is therefore defined as
\begin{equation}
    l_i=-\log p_{m_i}.
    \label{loss_per_sample_unsecure}
\end{equation}
In Section~\ref{security_structure_rl}, we will present a novel re-formulation of this per sample loss, which will enable the learning of the security enforcing structure.

\subsection{Policy Gradient Method for Wireless Communications}
Since the encoder is deterministic conditioned on a specific message $m$, one needs to introduce a perturbation on the codeword. This is usually done by an additive Gaussian noise $\mathbf{w}\sim \mathcal{N}(\mathbf{0},\mathbf{I}\sigma_{\pi}^2)$. Moreover, the codeword gets scaled such that the perturbed codeword $\mathbf{x}_p$ still obeys the power normalization. Therefore, we have that $\mathbf{x}_p= \sqrt{1-\sigma_{\pi}^2}\mathbf{x}_{\theta}+\mathbf{w}$. With this definition, the policy $\pi(\mathbf{x}_p|m)$ is given as (see also \cite{Aoudia-RL})
\begin{equation}
   \pi_{\theta}(\mathbf{x}_{p,i}|m_i) = \frac{1}{(\pi \sigma_{\pi}^2)^n}\exp \left(-\frac{\|\mathbf{x}_{p,i} - \sqrt{1-\sigma_{\pi}^2}\mathbf{x}_{\theta}\|}{\sigma_{\pi}^2} \right)
\end{equation} where $\mathbf{x}_{p,i}$ is the evaluation of the function $\mathbf{x}_p$ for the $i$-th sample of the message $m_i$. We therefore have 
\begin{equation}
  \log \pi_{\theta}(\mathbf{x}_{p,i}|m_i) = -\tfrac{1}{\sigma_{\pi}^2}\|\mathbf{x}_{p,i}-\sqrt{1-\sigma_{\pi}^2}\mathbf{x}_{\theta}\|+c.
\end{equation}
For the training of the encoder weights $\theta$, one can now feed 
\begin{equation}
    \nabla_{\theta}J(\theta)\approx \frac{1}{k}\sum_i^{k} l_i \nabla_{\theta}\log \pi_{\theta}(\mathbf{x}_{p,i}|m_i)
    \label{J_per_sample}
\end{equation} to an optimizer like Nadam\cite{Nadam} and train the NN.

\subsection{Enforcing Structure and Security Constraints on the Encoder}
\label{security_structure_rl}

As argued above in Section \ref{InfoThModel}, due to the major challenges of incorporating a secrecy criterion based on information theoretic metrics, we opt for a secrecy constraint based on the cross-entropy metric. We therefore take a similar approach as in \cite{ICC_DL_Sec}. A cross-entropy based metric is also dependent on a decoder, which is why we need to introduce a second decoder, which enforces a particular structure. From now on we refer to this as the structure enforcing (SE) decoder. This decoder needs to share the noise parameter with Eve, to apply the secrecy methods from \cite{ICC_DL_Sec} to our case. With this decoder, we can enforce a co-set-like structure on the resulting constellation. There, the data-carrying messages label the co-set and the particular codeword inside the co-set/cluster are chosen at random. This technique therefore mimics classical co-set coding methods, which go back to the seminal work of Wyner \cite{Wyner75}. We further refer the reader to \cite[Appendix A]{Oggier16} for further discussions on how co-sets can enable secrecy. Intuitively, the eavesdropper can only distinguish between clusters of codewords, but not between the codewords inside each cluster. The legitimate receiver Bob however, has a better channel and can use his advantage to also distinguish between the codewords inside the cluster. Our objective is therefore to produce a clustered constellation from the cross-entropy loss. This constellation can then be used for secure encoding afterwards. To enable this cluster structure in our NN encoding, we introduce a cross-entropy loss constraint for our SE decoder which is fed with a modified input distribution. This approach follows previous work in \cite{ICC_DL_Sec}, and for convenience and completeness we will repeat the basic construction here.

The goal of the modification is to obtain clusters of codewords (calculated with the $k$-means method) that have the same input probability. Normally, due to the one-hot encoding approach, a certain symbol has probability one if it was sent in the sample in the batch. Consider for example the training vector batch $\mathbf{m}=(1,2,3,4)$, resulting in the one-hot data matrix
\begin{equation*}
\mathbf{S}=
  \begin{bmatrix}
    1 & 0 & 0 & 0 \\
    0 & 1 & 0 & 0 \\
    0 & 0 & 1 & 0 \\
    0 & 0 & 0 & 1 \\
  \end{bmatrix}\end{equation*} where the rows are the samples of the batch and the columns indicate the symbol. We now modify the true data matrix towards an equalized matrix $\bar{\mathbf{S}}$:
\begin{equation*}
 \bar{\mathbf{S}}=\mathbf{S}\mathbf{E}=
  \begin{bmatrix}
    0.5 & 0.5 & 0 & 0 \\
    0.5 & 0.5 & 0 & 0 \\
    0 & 0 & 0.5 & 0.5 \\
    0 & 0 & 0.5 & 0.5\\
  \end{bmatrix}
  \end{equation*} where, for example, the first sample has an equal probability to be symbol $1$ and symbol $2$. The matrix $\mathbf{E}$ can be calculated with the $k$-means algorithm in conjunction with Algorithm $1$ from \cite{ICC_DL_Sec}.
The SE decoder's cross-entropy can now be written as   
  \begin{IEEEeqnarray*}{rCl}
H(p_{\overline{\text{data}}}(M)),p_{\text{SE}}(M))=- \sum_{m\in\mathcal{M}} \bar{s}_m \log \tilde{s}_m,
\label{Loss_Eve}
\end{IEEEeqnarray*}
 where the vectors $\tilde{\mathbf{s}}$ and $\bar{\mathbf{s}}$ can be interpreted as the decoded distribution and as the equalized input symbol distribution, respectively, as both are normalized to one. Averaging over $k$ samples yields the loss
 \begin{equation}
L_{\text{SE}}=-\frac{1}{k}\sum_{i=1}^k  \sum_{m\in\mathcal{M}} \bar{s}_{m} \log \tilde{s}_{m,i},
\label{Loss_Eve_2}
 \end{equation} which has a different form than \eqref{Loss_unsecure}. This is due to the fact that in \eqref{Loss_unsecure}, we used one-hot encoded messages, which pick the corresponding $\log p_m$ term from the sum, and set the others to zero. However, in our secure encoding scenario, we have the equalized vector $\bar{s}$, which picks all $\log \tilde{s}_m$ terms, that are uniformly distributed with $p>0$ in their cluster. Therefore, strictly speaking, the sum in \eqref{Loss_Eve_2} is now only over clusters/co-sets.
 
Now, in this paper we want to enable secrecy by applying RL via the policy gradient method. This means that we need a per sample loss. Interestingly, \eqref{Loss_Eve_2} is in a form such that we can extract a per samples loss as follows:
 \begin{equation*}
     l_{\text{SE},i}:= -\sum_{m\in\mathcal{M}} \bar{s}_m \log \tilde{s}_{m,i}.
\label{Per_sample_loss_Eve}
 \end{equation*}
 This loss can be seen as a per sample secrecy constraint which takes into account the whole cluster around a specific sample. Therefore, in the exploration step, the whole cluster influences the decision process. Together with the previous encoder per sample loss function, we can define a new security enabled per sample loss function 
 \begin{equation}
    l_{\text{sec},i}=(1-\alpha) l_i-\alpha l_{\text{SE},i},
 \end{equation} where $\alpha\in [0,1]$ controls the influence of the secrecy structure enforcing constraint. Therefore, the parameter $\alpha$ controls the trade-off between security and communication rate on the legitimate channel.
We can now re-formulate the RL-objective, i.e., the gradient of $J$ in \eqref{J_per_sample}, such that it is a security-enabled gradient update as 
\begin{equation}
    \nabla_{\theta}J_{\text{sec}}(\theta)\approx \frac{1}{k}\sum_i^{k} l_{\text{sec},i} \nabla_{\theta}\log \pi_{\theta}(\mathbf{x}_{p,i}|m_i).
    \label{J_per_sample_sec}
\end{equation}  
 The new security structure enforcing training algorithm, which uses the policy gradient method, is shown in Algorithm~$1$.

\begin{figure}
 \removelatexerror
{
\begin{algorithm}[H]
    \LinesNumberedHidden
    \SetKwInput{Input}{Require}
    \SetKwInput{Output}{Output}
    \SetKwInput{Initialize}{Initialize}
	\Input{SE Decoder, \emph{Equalization operator }$\mathbf{E}\in\mathbb{R}^{|\mathcal{M}|\times  |\mathcal{M}|}$}
	\While{stopping criterion not met}{
	\textbf{Train Decoder:}
	
	\Indp
	
    \Input{encoder with randomly initialized weights $\theta$}
    \Initialize{sample batch from source $\mathbf{m}_{\text{oh}}$}
        $\mathbf{X}(\mathbf{M}_{\text{oh}}) \leftarrow \text{encoder}$ \\
        
        $\mathbf{Y} \leftarrow \text{channel }  P(Y|X)$ \\
        $\hat{\mathbf{M}}(\mathbf{M}_{\text{oh}}) \leftarrow \text{decoder with weights } \psi$ \\
        $L(\hat{\mathbf{M}},\mathbf{M}_{\text{oh}}) \leftarrow \text{cross entropy loss}$ \\
        
        $\psi \leftarrow \text{Nadam optimzer on } L(\psi)$
        
    \Indm
    
    \textbf{Train Encoder:}
    
    \Indp
         \Input{encoder with weights $\theta$}
         \Input{structure enforcing (SE) decoder}
         \Input{decoder with learned weights $\psi$}
        \Initialize{sample batch with size $k$ from one-hot source $\mathbf{m}_{\text{oh}}$}
        \Initialize{Equalization operator $\mathbf{E}$}
        
        $\mathbf{X}(\mathbf{M}_{\text{oh}}) \leftarrow \text{encoder with weights } \theta$ \\
        $\mathbf{X}_p \leftarrow \text{policy: } \sqrt{1-\sigma_p}\mathbf{X}+\mathbf{W}$
        
        $\mathbf{Y} \leftarrow \text{channel }  P(Y|X_p)$ \\
        $\mathbf{Z} \leftarrow \text{channel }  P(Z|X_p)$ \\
        $\hat{\mathbf{M}}(\mathbf{M}_{\text{oh}}) \leftarrow \text{decoder with weights }\psi$ \\
        $\hat{\mathbf{M}}_{\text{SE}}(\mathbf{M}_{\text{oh}}) \leftarrow \text{structure enforcing decoder}$ \\
        $l_i(\hat{\mathbf{m}},\mathbf{m}_{\text{oh}}) \leftarrow \text{per sample loss}$ \\
        $\mathbf{M}_{\text{eq}}\leftarrow \mathbf{E}\mathbf{M}_{\text{oh}} $\\
        $l_{i,\text{SE}}(\hat{\mathbf{m}}_{\text{SE}},\mathbf{m}_{\text{eq}}) \leftarrow \text{new per sample loss}$ \\
        $l_{\text{sec},i} \leftarrow (1-\alpha)l_i+\alpha l_{\text{SE},i}$ \\
        $\log \pi_{\theta}(\mathbf{x}_{p,i}|m_i) \leftarrow -\frac{||\mathbf{x}_{p,i} - \mathbf{x}\sqrt{1-\sigma_p}||^2}{\sigma_p^2}$ \\
        $J(\theta) \leftarrow \tfrac{1}{k}\sum_i l_{\text{sec},i}\nabla_{\theta}\log \pi_{\theta}(\mathbf{x}_{p,i}|m_i)$ \\
        $\theta \leftarrow \text{Nadam optimzer on } J(\theta)$
        
    \Indm

    }
%\caption{Data pmf equalization operator}
    \caption{This algorithm trains the decoder and encoder alternatingly. The decoder is trained with a standard procedure, while the encoder is trained via a policy gradient method. Moreover, we enable secure encoder training with a modified per sample loss.}
    \label{algo:label_alternate_sec_train}
\end{algorithm}
}
%\vspace{-1em}
\end{figure}

\subsection{Training of the Encoder-Decoder Network}

\subsubsection{Encoder-decoder training without secrecy structure}
To train the encoder, we use a signal-to-noise ratio (SNR) per bit of $E_{\text{b}}/N_0=7$ dB. This specifies the noise variance of the direct intended channel in relation to our normalized codeword power. When we add the perturbation, the codewords are scaled such that they are still normalized. Moreover, we assume an SNR per bit of $E_{\text{b}}/N_0 = 6$ dB for Eves channel, which corresponds approximately to an $E_{\text{b}}/N_0=12$ dB additional noise factor on top of Bob's channel. The training of the encoder-decoder pair of Alice and Bob, before adjusting for security, is done similar to \cite{Aoudia-RL}. This means we use an alternating optimization, where we start with optimizing the decoder weights $\psi$ of Bob with randomly initialized encoder weights $\theta$ of Alice and the usual cross-entropy loss metric, together with the Nadam optimizer. This does not require channel knowledge or policy gradient methods, since we only need the gradient of the decoder. After that we train for the encoder weights $\theta$, with the policy gradient method, to optimize $\theta$ with Nadam, providing the gradient in \eqref{J_per_sample} for the update, without secrecy constraint. This is done iterative for $2$ epochs with $400$ steps, where each step draws a new batch of messages.
\subsubsection{Encoder-decoder training with secrecy structure}
For the security constraint, we need the SE decoder, which is implemented with a standard NN decoder with a hidden layer with elu activation and an output layer with softmax activation. The decoder will be pre-trained with a batch size of $200$ and $400$ iterations per epoch, for $4$ epochs with a learning rate of $0.005$ with the Nadam optimizer. Afterwards, we initialize the matrix $\mathbf{E}$ and train the Alice-Bob encoder-decoder pair, which will be optimized for secure encoding with Alg.~\ref{algo:label_alternate_sec_train} and the new per sample loss \eqref{J_per_sample_sec} over $2$ epochs, a batch size of $500$, with $400$ iterations, a learning rate of $0.005$ and $\alpha = 0.7$. The simulation code is available at \cite{Fritschek2020_code}, implemented with TensorFlow 2 \cite{tensorflow2015-whitepaper}.

\section{Evaluation}

\begin{figure}
\vspace{0.5em}
\centering
\includegraphics[scale = 0.48]{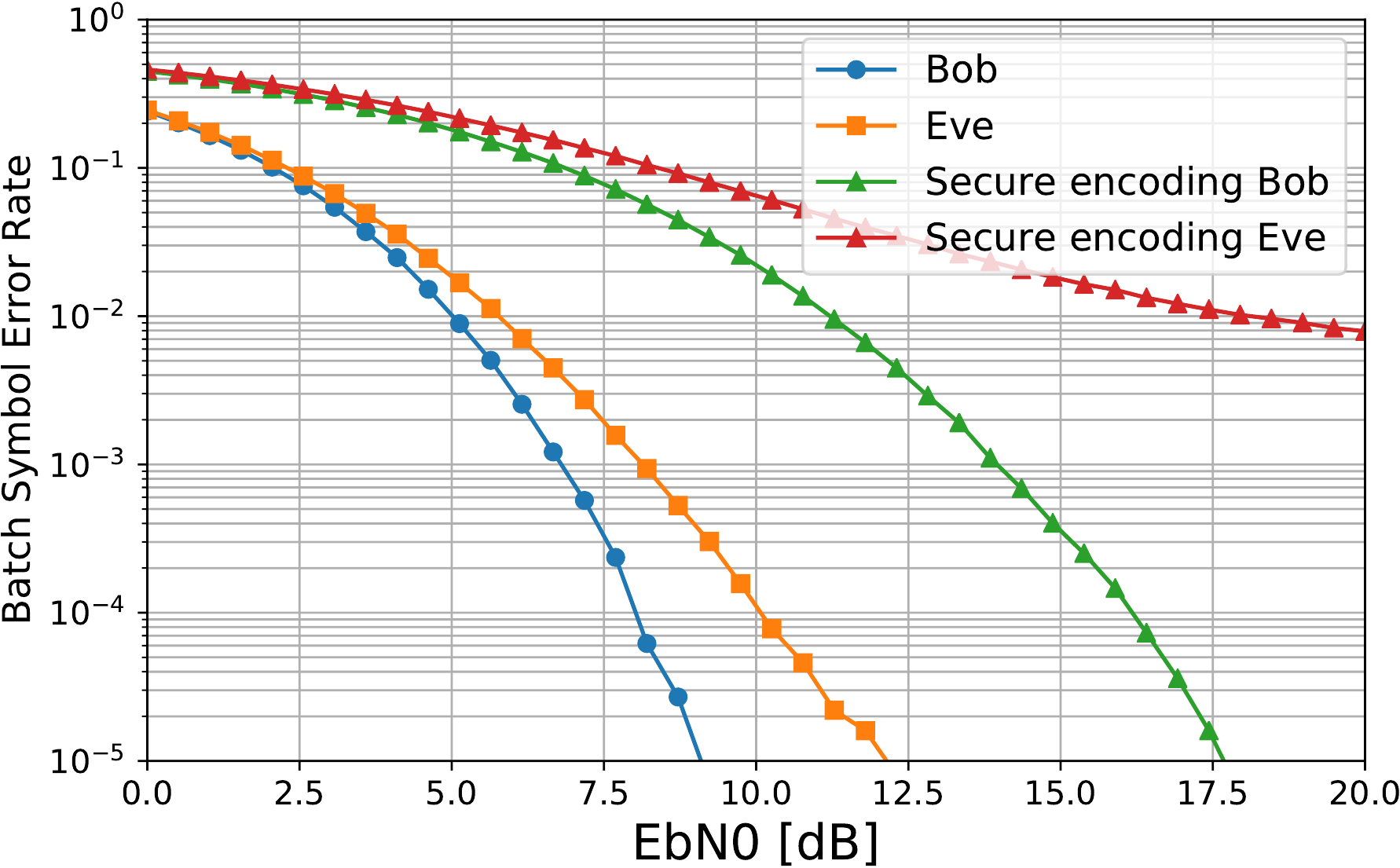}
\caption{Evaluation of the proposed method for a $32$-dimensional codeword constellation. Bob and Eve show the error rate for transmission of the codewords before secure encoding and, accordingly, Secure encoding Bob and Secure encoding Eve refer to after secure encoding.}
\label{fig:res}
\end{figure}

For the evaluation we use NN decoders with the standard cross-entropy loss function. Moreover, Bob and Eve use the same decoder structure to have a fair comparison. Both decoders use one hidden layer with elu activation which yields better results compared to Relu in our simulations.
After the alternating training of Alice and Bob, and the training of the SE decoder, and the secrecy-enabled alternating training of Alice and Bob, we have an encoding system for secure communication. We now train the decoder Bob one more time and train another decoder as a representative for Eve. Both are trained with the same parameters for 400 iterations, and a batch size of 200. We test the system once before, and once after training for secure encoding, with $10^6$ samples for each $E_{\text{b}}/N_0$ data point. We model Eve's channel with an additive fixed noise of $E_{\text{b}}/N_0=12$ dB which is additionally to Bob's noise and helps to unify both results in one figure.
Note that after secure training of the encoder, we utilize the resulting co-set structure and use a message set with $4$ messages, then we randomly choose the satellite codeword inside the cluster, mapping the $4$ symbols code to a $16$ symbols code. The evaluation results in Fig.~\ref{fig:res} show that Bob's and Eve's performance is similar good with a relatively low symbol error rate per batch. In these results, the error rate for Eve's signal is worse than Bob's due to higher baseline noise of Eve's received signal. After we have trained the system for secure encoding, both error rates are elevated. Bob's error rate is higher due to the trade-off between security and communication rate, but still declines with a higher $E_{\text{b}}/N_0$ ratio. On the other hand, Eve's error rate is elevated but stays flat above a certain threshold, even for high $E_{\text{b}}/N_0$ values. This shows that security can be achieved in this scenario.

%\begin{figure}[htb]
%
%\begin{minipage}[b]{.48\linewidth}
%  \centering
%  \centerline{\includegraphics[width=4cm]{const_encoding-crop.pdf}}
%  \vspace{1.5cm}
%  \centerline{(b) Standard constellation}\medskip
%\end{minipage}
%\hfill
%\begin{minipage}[b]{0.48\linewidth}
%  \centering
%  \centerline{\includegraphics[width=4cm]{const_secure_encoding-crop.pdf}}
%  \vspace{1.5cm}
%  \centerline{(c) Secure constellation }\medskip
%\end{minipage}
%
%\caption{Constellations before and after secure encoding for $2$ dimensions and $16$ constellation points.}
%\label{fig:res_2d}
%
%\end{figure}

\begin{figure}
    
	\centering
	\subfigure[][Standard constellation]{\label{fig:sym1}%
		\includegraphics[width=4cm]{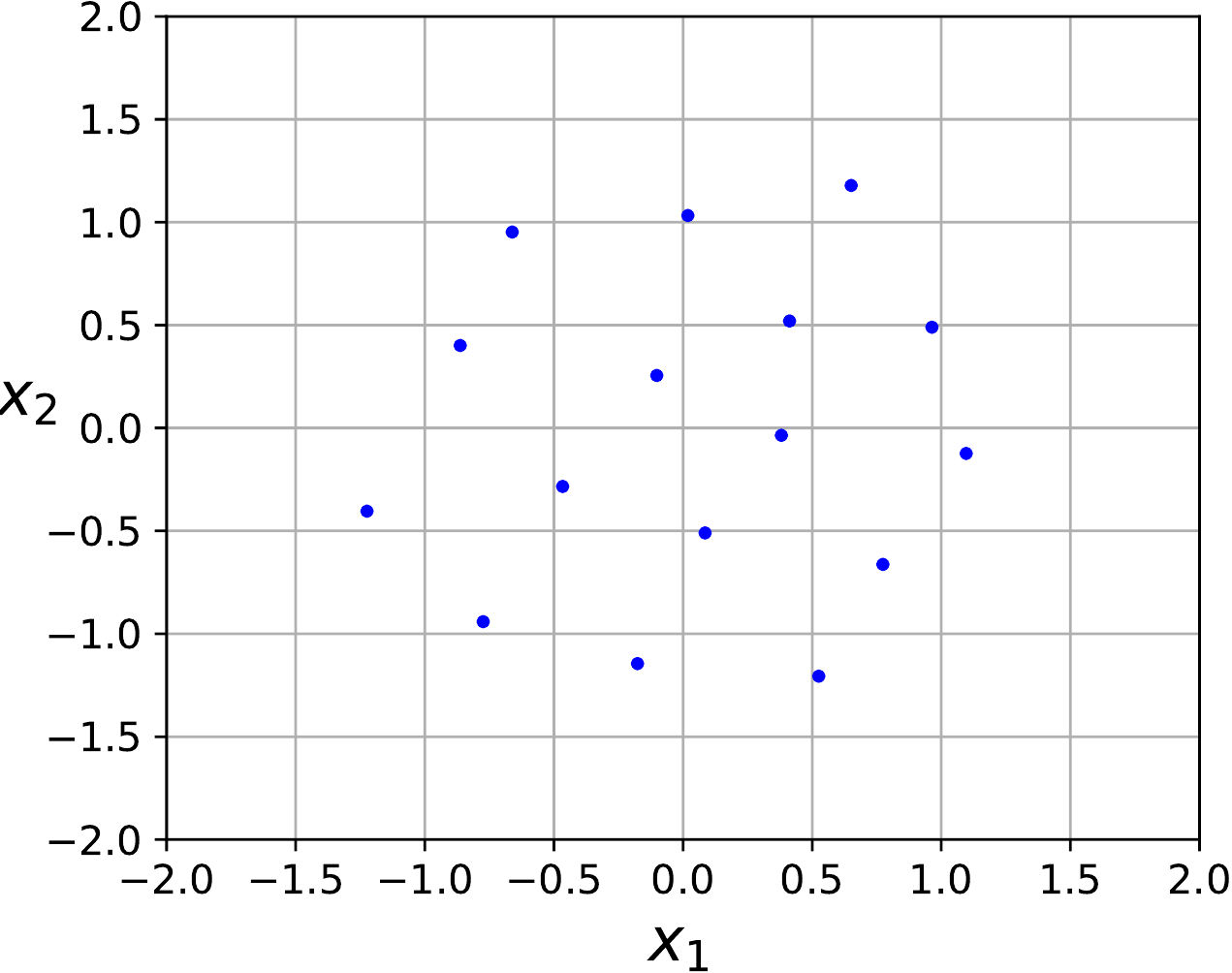}}
	\hspace{0.5cm}
	\subfigure[][Secure constellation]{\label{fig:sym2}%
		\includegraphics[width=4cm]{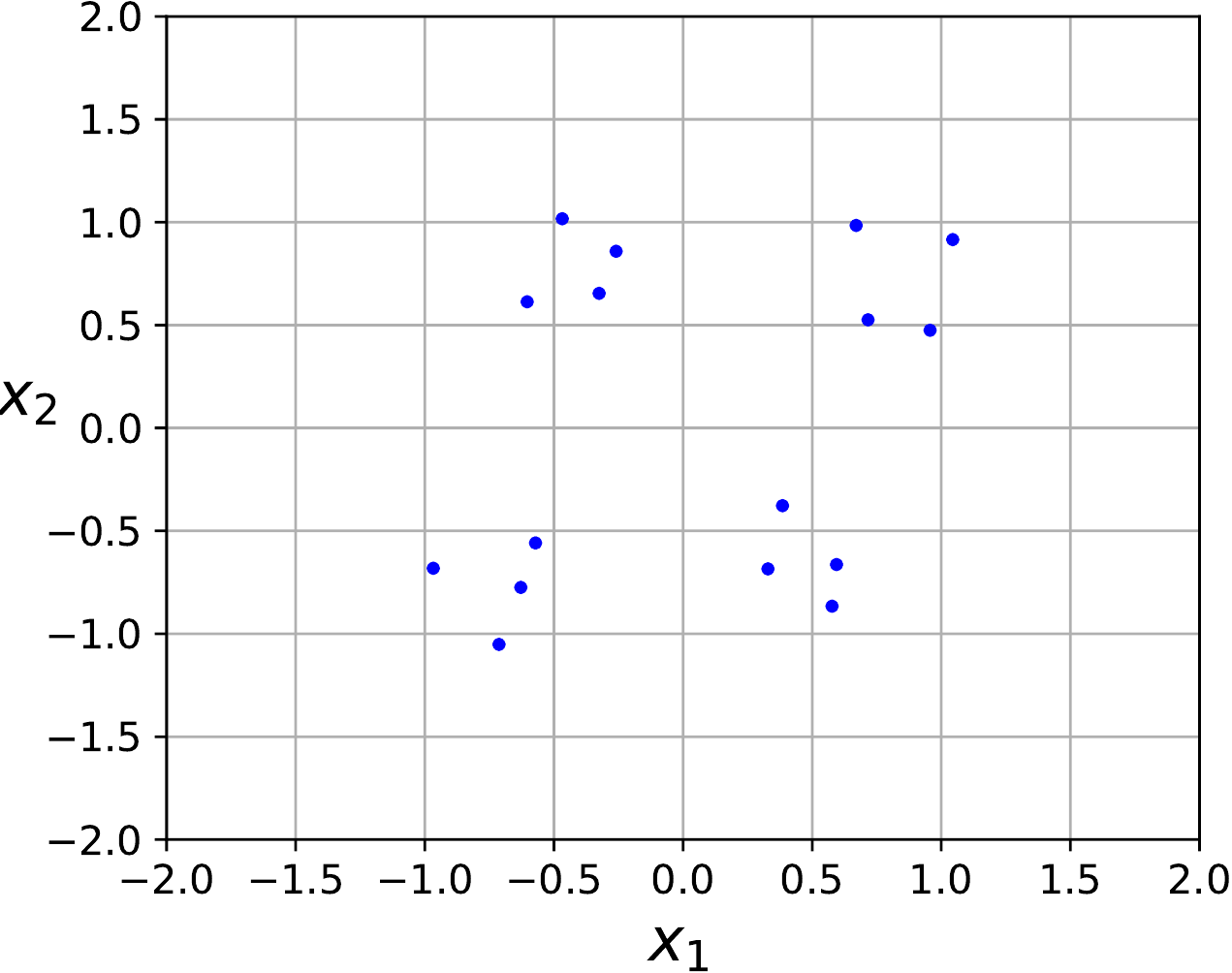}}
	\caption{Constellations before and after secure encoding for two dimensions and $16$ constellation points.}
\label{fig:res_2d}
\end{figure}

Furthermore, we trained the NN models for two signal transmission dimensions, to visualize the constellations. We note that one could use t-SNE on the higher dimensional signals from above, however, this output would be highly dependent on the used parameters.
The resulting constellations before and after our secure encoding process can be seen in Fig.~\ref{fig:res_2d}. We note that the training iterations were lowered and we used early stopping after $260$ iterations in the second epoch. Moreover, we tuned the security trade-off parameter to $\alpha=0.5$. All other parameters are the same as in the $32$ dimensional case. Here, it can be seen that the system forms clusters and that the structure enforcing method works as intended.

\section{Conclusions}
We have shown that a recently proposed model-free approach for training NN encoder-decoder pairs for reliable communication, using policy gradient methods, can be extended to produce a secrecy enforcing modulation structure. The challenge was to define a meaningful per sample loss that obeys constraints for secrecy, as well as works for policy gradient methods. For that we re-visited another recent approach which enabled co-set structures for NN encoders with a modified input function for regular model-aware training. This approach was based on one-hot encoded messages in conjunction with the regular cross-entropy loss.  We showed how to extract a per sample loss from this method, which conserves its structure enforcing properties. This makes it possible to use in conjunction with the policy gradient method. In our simulations, we showed that the security enhanced policy gradient exploration, with our novel per sample loss, can indeed enable an advantage in terms of error for Bob, and therefore secure communication.
Moreover, we showed that the learned modulation indeed produces cluster structures, which enable the secure communication.
\label{sec:conclusions}

%\pagebreak
\bibliographystyle{IEEEbib}
\bibliography{strings,refs}

\end{document}